\documentstyle[epsf,graphics,epsfig]{mn}

\newcommand{\etal}{{et~al.}}

\def\hompc{\,h\,{\rm Mpc}^{-1}}
\def\mpcoh{\,h^{-1}\,{\rm Mpc}}
\def\bj{b_{\scriptscriptstyle\rm J}}
\def\spose#1{\hbox to 0pt{#1\hss}}

\def\simgt{\mathrel{\lower0.6ex\hbox{$\buildrel {\textstyle >}
 \over {\scriptstyle \sim}$}}}
\def\simlt{\mathrel{\lower0.6ex\hbox{$\buildrel {\textstyle <}
 \over {\scriptstyle \sim}$}}}

\def\m@th{\mathsurround=0pt }
\def\eqalign#1{\null\,\vcenter{\openup1\jot \m@th
 \ialign{\strut\hfil$\displaystyle{##}$&$\displaystyle{{}##}$\hfil
 \crcr#1\crcr}}\,}
\def\japitem#1{\smallskip\noindent\rlap{#1}\hglue\parindent\hangindent\parindent}
\def\enditem{\smallskip\noindent}

\title[2dFGRS: Power Spectrum] {The 2dF Galaxy Redshift Survey: The
power spectrum and the matter content of the universe} 
\author[Will J.\ Percival \etal]{
\parbox[t]{\textwidth}{
Will J.\ Percival$^1$,
Carlton M.\ Baugh$^2$,
Joss Bland-Hawthorn$^3$,
Terry Bridges$^3$, 
Russell Cannon$^3$, 
Shaun Cole$^2$, 
Matthew Colless$^4$, 
Chris Collins$^5$, 
Warrick Couch$^6$, 
Gavin Dalton$^7$,
Roberto De Propris$^6$,
Simon P.\ Driver$^8$, 
George Efstathiou$^9$, 
Richard S.\ Ellis$^{10}$, 
Carlos S.\ Frenk$^2$, 
Karl Glazebrook$^{11}$, 
Carole Jackson$^4$,
Ofer Lahav$^9$, 
Ian Lewis$^3$, 
Stuart Lumsden$^{12}$, 
Steve Maddox$^{13}$,
Stephen Moody$^9$,
Peder Norberg$^2$,
John A.\ Peacock$^1$,
Bruce A.\ Peterson$^4$, 
Will Sutherland$^1$,
Keith Taylor$^3$}
\vspace*{6pt} \\ 
$^1$Institute for Astronomy, University of Edinburgh, Royal Observatory, 
       Blackford Hill, Edinburgh EH9 3HJ, UK \\
$^2$Department of Physics, University of Durham, South Road, Durham DH1 3LE, UK \\
$^3$Anglo-Australian Observatory, P.O.\ Box 296, Epping, NSW 2121,
    Australia\\  
$^4$Research School of Astronomy \& Astrophysics, The Australian 
    National University, Weston Creek, ACT 2611, Australia \\
$^5$Astrophysics Research Institute, Liverpool John Moores University,  
    Twelve Quays House, Birkenhead, L14 1LD, UK \\
$^6$Department of Astrophysics, University of New South Wales, Sydney, 
    NSW 2052, Australia \\
$^7$Department of Physics, University of Oxford, Keble Road, Oxford OX1 3RH, UK \\
$^8$School of Physics and Astronomy, University of St Andrews, North
    Haugh, St Andrews, Fife, KY6 9SS, UK \\
$^9$Institute of Astronomy, University of Cambridge, Madingley Road,
    Cambridge CB3 0HA, UK \\
$^{10}$Department of Astronomy, Caltech, Pasadena, CA 91125, USA \\
$^{11}$Department of Physics \& Astronomy, Johns Hopkins University,
       Baltimore, MD 21218-2686, USA \\
$^{12}$Department of Physics, University of Leeds, Woodhouse Lane,
       Leeds, LS2 9JT, UK \\
$^{13}$School of Physics \& Astronomy, University of Nottingham,
       Nottingham NG7 2RD, UK \\
}

\date{Submitted for publication in MNRAS}
\begin{document}
\maketitle

\begin{abstract}
The 2dF Galaxy Redshift Survey has now measured in excess of
$160\,000$ galaxy redshifts. This paper presents the power spectrum of
the galaxy distribution, calculated using a direct FFT-based
technique. We argue that, within the $k$-space region $0.02 \simlt k
\simlt 0.15\hompc$, the shape of this spectrum should be close to that
of the linear density perturbations convolved with the window function
of the survey. This window function and its convolving effect on the
power spectrum estimate are analyzed in detail. By convolving model
spectra, we are able to fit the power-spectrum data and provide a
measure of the matter content of the universe. Our results show that
models containing baryon oscillations are mildly preferred over
featureless power spectra. Analysis of the data yields 68\% confidence
limits on the total matter density times the Hubble parameter
$\Omega_m h = 0.20 \pm 0.03$, and the baryon fraction
$\Omega_b/\Omega_m = 0.15 \pm 0.07$, assuming scale-invariant
primordial fluctuations.
\end{abstract}

\section{introduction}

Present-day cosmological structure is thought to have formed by the
gravitational amplification of small density perturbations. These
fluctuations are readily quantified in terms of their Fourier modes
via the power spectrum, which is a statistically complete description
for a Gaussian field.  The power spectrum is also of direct physical
interest, because it encodes information about the formation of the
primordial fluctuations, and especially about how these are modified
according to the matter content of the universe.

In this paper, we present an estimate of the power spectrum of the
galaxy distribution in the 2dF Galaxy Redshift Survey (2dFGRS).  The
2dFGRS is designed around the 2dF multi-fibre spectrograph on the
Anglo-Australian Telescope, which is capable of obtaining spectra for
up to 400 objects simultaneously over a 2 degree diameter field of
view. Full details of the instrument and its performance are given in
Lewis \etal\ (2000). See also {\tt http://www.aao.gov.au/2dF/}.  The
survey aims to obtain redshifts for $250\,000$ galaxies to an
extinction-corrected magnitude limit of $\bj < 19.45$.  A description
of the survey, with preliminary results, is given by Colless (1999);
full details of the present status can be obtained from {\tt
http://www.mso.anu.edu.au/2dFGRS/}.

At the time of writing, the 2dFGRS is the largest existing galaxy
redshift survey, following a natural progression from studies such as
the CfA survey (Huchra \etal\ 1990), the LCRS (Shectman \etal\ 1996),
and the PSCz survey (Saunders \etal\ 2000). The data and analysis
presented in this paper covers the sample with $166\,490$ redshifts
observed prior to February 2001. A sample of this size allows
large-scale structure statistics to be measured with very small random
errors, and we present an initial power-spectrum analysis of the
2dFGRS here. Section 2 details some of the practical issues concerning
sample selection, and Section 3 discusses power-spectrum estimation.
The survey coverage in angular position and redshift is relatively
complex, and the convolving effects of the survey window are
significant compared to the small random errors. These effects are
therefore studied in some detail, both analytically and in comparison
to mock data, in Section 4. This leads to a robust estimate of the
covariance matrix for the estimates of the power at different
wavenumbers, which is presented in Section 5.  The covariance matrix
allows proper likelihood-based model fitting, which is carried out in
Section 6.  The power-spectrum fits clearly indicate a low-density
universe with $\Omega_m h\simeq 0.2$, in agreement with many past
studies. We also show that the preferred model requires a degree of
baryon oscillations in the power spectrum, corresponding to a baryonic
fraction of about 15\%.  We conclude by considering the consistency
between this picture and other lines of evidence.

\section{the 2dFGRS sample}  \label{sec:2dFGRS_sample}

\subsection{The angular mask}  \label{sec:angular_mask}

When complete, the angular geometry of the 2dFGRS will consist of two
declination strips plus 100 random 2-degree fields. One strip is near
the Southern Galactic Pole (SGP) and covers approximately
85$^\circ$$\times$15$^\circ$; the other strip is near the Northern
Galactic Pole (NGP) and covers 75$^\circ$$\times$10$^\circ$. These
strips are not coplanar, which is a significant factor in using the
survey to measure 3D structure. The 100 random fields are spread
uniformly over the 7000~deg$^2$ region of the APM catalogue near the
SGP; the present analysis includes 71 of these fields.

The input catalogue is a revised and extended version of the APM
galaxy catalogue (Maddox \etal\ 1990a,b,c,1996). This includes over 5
million galaxies down to $\bj=20.5$ over $\sim10^4$\,deg$^2$.  The APM
catalogue was used previously to recover the 3D power spectrum of
galaxies by inverting the appropriate integral equations (Baugh \&
Efstathiou 1993; Efstathiou \& Moody 2000). However, these techniques
are demanding in sample variance and photometric uniformity, and we
expect that a fully 3D analysis should yield a more robust result.

An adaptive tiling algorithm is employed to cover the survey area with
a minimum number of overlapping 2dF fields. With this algorithm we are
able to achieve a 93\% sampling rate with on average fewer than 5\%
wasted fibres per field. Over the whole area of the survey there are
in excess of $250\,000$ galaxies.  At the present intermediate stage
of the survey, unobserved fields mean that the proportion of targets
with known redshifts is a strongly varying function of position.  In
addition, regions around bright stars are omitted, so the 2dFGRS
angular mask is a complicated pattern on the sky (see e.g. Colless
1999). Nevertheless, because the tiling algorithm is known, it is
possible to generate random catalogues that are subject to the same
selection effects. A number of different codes have been written to
achieve this task, with consistent results. Furthermore, because a 3D
power spectrum analysis averages over directions, small imperfections
in reproducing the sky pattern of the real data tend to wash out. For
example, we tried adding magnitude offset errors of $\Delta M=\pm0.2$
in each $5^\circ$ Schmidt field, but the power spectrum did not change
significantly.

Given the sampling pattern on the sky, there are two possible analysis
strategies: one can either build a similar variation into any random
catalogue, or the analysis can use a uniform random catalogue,
weighting each galaxy by the reciprocal of the sampling. The former
strategy is superior in terms of shot noise, but the latter is
necessary if the mask is correlated with real structure (e.g. fibre
crowding problems in high-density regions). We obtain almost identical
results with either strategy, demonstrating that the adaptive tiling
has achieved its target of uniform selection of targets.

\subsection{Redshift selection}  \label{sec:nz}

The sample is chosen to be magnitude-limited at $\bj=19.45$ after
extinction-correcting all the magnitudes in the APM catalogue
(Schlegel, Finkbeiner \& Davis 1998). This limit was chosen because
the mean number of galaxies per square degree then matches the density
of fibres available with 2dF. The resulting distribution of galaxy
redshifts $n(z)\,dz$ has a median of approximately 0.11, and can be
fitted by
\begin{equation}
  n(z)\,dz\propto\frac{\left(z / z_c\right)^{\gamma-1}}
    {\left[1+\left(z/z_c\right)^{\gamma/\beta}\right]^{1+\beta}}\;dz,
    \label{eq:nz}
\end{equation}
where $z_c$, $\gamma$ \& $\beta$ are fitted parameters. Fitting to all
of the galaxy redshifts gives $z_c=0.144$, $\gamma=2.21$ \&
$\beta=0.554$. However, the redshift distribution is expected to vary
with position on the sky, because the survey depth is not completely
uniform. This arises because the spectroscopic success rate is a
function of apparent magnitude: data from poorer nights are biased to
brighter objects, and thus to lower redshifts. Also, our estimates of
galactic extinction and CCD calibration of the zero points of the
individual photographic plates have been revised since the original
input catalogue was defined. All these effects contribute to a
modulation of the depth of the survey, which is accounted for when
making the random catalogue that defines the survey volume. Because
these estimates of nonuniformity can never be quite precise, we have
chosen to allow the parameters of the $n(z)$ fit to be different in
distinct zones of the sky, treating the NGP, SGP, and random fields
separately. Analysis of mock catalogues shows that this makes only a
small difference to the power estimates at $k>0.02\hompc$, which we
use as our normal analysis limit.

\section{estimating the power spectrum}

\subsection{The FKP method}  \label{sec:FKP_method}

Following the FFT-based approach of Feldman, Kaiser \& Peacock (1994;
FKP), each galaxy in the sample was placed onto a
$512\times512\times256$ grid, scaled to cover the entire sample in
each direction. The size of the cuboid that just contains all of the
data to a redshift limit at $z=0.25$ is $1388\times1017\times685\,
h^{-3}{\rm Mpc}^3$, assuming a flat $\Omega_m=0.3$ cosmology. The grid
used therefore leads to approximately equal Nyquist frequencies in
each direction. For this grid, $k=0.15\hompc$\ corresponds to a
conservative $13\%$ of the minimum Nyquist frequency. Comparing with
results obtained using different size grids also suggests that
aliasing is not a problem for frequencies $k<0.15$, although, of
course, it will affect the result at large $k$.  As shown by FKP, the
galaxies must be weighted in order to optimize the balance between
cosmic variance and shot noise. Assuming that the fluctuations are
Gaussian, the optimal FKP weight is
\begin{equation}
  w(z,\theta,\phi)=\frac{1}{\left[1+\bar{P} \bar{n}(z,\theta,\phi)\right]},
\end{equation}
where $\bar{n}(z,\theta,\phi)$ is the expected galaxy density. A value
of $\bar{P} = 5000\,h^3\,\rm Mpc^{-3}$ was assumed, but this is not
critical. For a high-density sample like the 2dFGRS, one very nearly
gives equal weight to each volume element, independent of $\bar{P}$.
We applied an upper redshift limit of $z=0.25$, to remove regions with
a very low galaxy density where the choice of $\bar{P}$ would
matter. We also defined a lower limit of $z>0.003$. With restriction
to objects with redshift quality flag $\ge3$, this leaves a sample of
$147\,024$ objects.

In order to transform from redshifts to distances in $\mpcoh$ we need
to assume values for $\Omega_m$ and $\Omega_v$. Strictly, when fitting
models to the data we should alter these values to match the
model. However, the power spectrum is only weakly dependent on this
choice: we have tried both an Einstein-de Sitter cosmology and a flat
$\Omega_m=0.3$ cosmology and find approximately the same best fit
model parameters (see Table~\ref{tab:fit_data}). Normally, we will
present results for a flat $\Omega_m=0.3$ cosmology.

Before Fourier transforming the data, we need to convert from a
distribution of galaxies to a distribution of overdensities. To do
this, the survey volume was defined by a random catalogue that mimics
the sampling of the original data. In order to create such a sample,
we have used the known angular position and completeness of each field
(described in Section~\ref{sec:angular_mask}), coupled with an
empirically determined fit to the redshift distribution of the 2dFGRS
catalogue (described in Section~\ref{sec:nz}). The variations in
observing conditions, and the corresponding spectroscopic success
rate, mean that there are in practice variations in the survey depth
with sky position. These are allowed for by appropriate perturbations
of the redshift distribution in the random catalogue, although this
again turns out to be an unimportant effect. The random catalogue we
created had 5 times the number of points in the galaxy catalogue.

Having Fourier transformed the resulting overdensity field, we simply
subtracted the shot noise contribution from the result, and
re-normalized $P({\bf k})$ in order to correct for a differing sample
volume and FFT volume, and to correct for the weighting (FKP equations
2.1.3 and 2.1.10).  $P({\bf k})$ was then spherically averaged over
$k$-space shells.

The power spectrum of the galaxy distribution needs to be interpreted
with care, as it is altered from that of the initial density
fluctuations (the linear power spectrum) by a number of effects:

\japitem{(1)}The collapse of structures. The mass no longer obeys the 
linear power spectrum on small scales after it has undergone
non-linear collapse.

\japitem{(2)}Galaxies are not expected to form a Poisson sampling of the
underlying mass fluctuations, and are biased with respect to this
distribution. 

\japitem{(3)}Random oscillations of galaxies within larger collapsed
objects causes `fingers of God' redshift space distortions that damp
estimates of the small-scale power.

\japitem{(4)}The infall of galaxies into concentrations of mass on
large scales (Kaiser 1987) enhances the observed power in the radial
direction, creating a large-scale redshift distortion.

\enditem 
Additionally, the recovered power spectrum estimated using the FKP
approach is the convolution of the galaxy power spectrum with the
`window function' $|W_k|^2$ of the survey ($W_k$ being the Fourier
transform of the product of the survey mask and redshift selection
function times $w(z)$; see FKP equation 2.1.7):
\begin{equation}
  \hat P_{\rm FKP}({\bf k}) \propto
    \int P_{\rm true}({\bf k}-{\bf q})|W_k({\bf q})|^2\,d^3{\bf q}.
\end{equation}
In Section~\ref{sec:window} we show how we may model the effect of the
window function, by convolving power spectra with an analytic fit to
the spherically averaged window function, and empirically by analysing
realizations of Gaussian density fields within the survey region. The
inverse problem of deconvolving the window function from the power
spectra is not attempted in this work.

\subsection{The window function}  \label{sec:window}

\begin{figure}
  \setlength{\epsfxsize}{\columnwidth}
  \centerline{\epsfbox{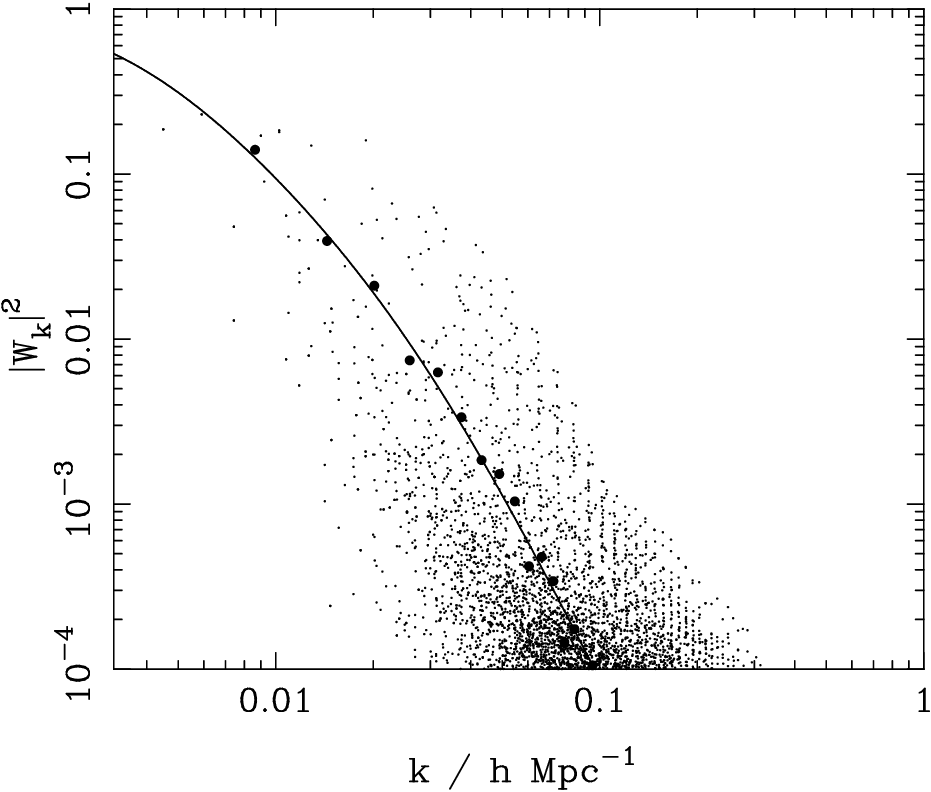}}

  \caption{The 2dFGRS window function in Fourier space. The fitting
  formula of Equation~\ref{eq:win_fit} is plotted (solid line)
  compared to the spherically averaged power obtained by Fourier
  transforming a random catalogue created to match the 2dFGRS window
  function (solid circles). This catalogue, containing 5 times as many
  points as the true galaxy catalogue was placed on a
  $512\times512\times256$ grid that encompassed all the data. Values
  from individual grid points are also plotted (dots), and show the
  anisotropy of the window function. }

  \label{fig:win}
\end{figure}

\begin{figure}
  \setlength{\epsfxsize}{\columnwidth} 
  \centerline{\epsfbox{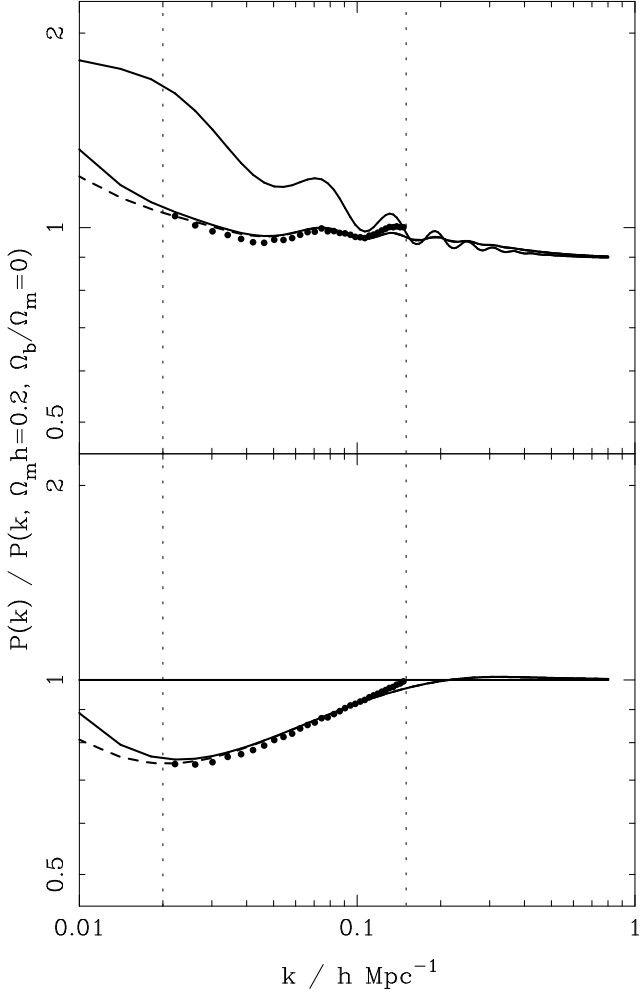}}

  \caption{The effect of convolving a linear power spectrum with the
  2dFGRS window function. Two model power spectra are compared: The
  top panel shows a power spectrum including a baryonic component
  $P(k,\Omega_mh=0.2,\Omega_b/\Omega_m=0.16)$ while the bottom panel
  shows a power spectrum $P(k,\Omega_mh=0.2,\Omega_b/\Omega_m=0)$ with
  no baryonic component. These spectra are divided by a smooth
  reference model, $P(k,\Omega_mh=0.2,\Omega_b/\Omega_m=0)$. The upper
  solid lines show the unconvolved power spectra, and the lower solid
  lines after numerically convolving with the fit to the window
  function given by Equation~\ref{eq:win_fit}. The dashed lines show
  the average power spectrum expected from data which has had the
  average over-density artificially set to zero (see
  Section~\ref{sec:window} for details). Solid symbols show the shape
  of the average recovered power spectrum from 1000 Gaussian
  realizations of the density field (with different phases), placed on
  a $256\times256\times128$ grid at locations covered by the
  2dFGRS. Due to the relatively coarse grid used, aliasing swamps the
  signal at $k>0.15\hompc$, and data are only shown for the range of
  $k$ over which the 2dFGRS power spectrum is fitted
  (Section~\ref{sec:fit}). This region is delineated by the vertical
  dotted lines.}

  \label{fig:win_pow}
\end{figure}

The 2dFGRS window function $|W_k|^2$ is not compact, and has a significant
effect on the recovered power spectrum. Although $W_k$ is anisotropic
due to the complicated real space selection function, we are
normally interested in the spherical average of the final power
spectrum over $k$-space shells. Therefore, if
$P_{\rm true}$ is isotropic, we obtain the same result by convolving
just with the spherical average of $|W_k|^2$.
A good approximation to the averaged window is
\begin{equation}
  \langle |W_k|^2 \rangle = [1 + (k/0.00342)^2 + (k/0.00983)^4]^{-1}.
    \label{eq:win_fit}
\end{equation}
This fit is compared to the exact 2dFGRS window in
Fig.~\ref{fig:win}. The potential cause of problems is the tail to
high $k$.  The convolution involves a normalization factor $\int | W_k
|^2 4\pi k^2 \: dk$, and a significant part of this volume lies
outside the `core' that has a scale of $0.003\hompc$. In this respect,
the 2dFGRS window is quite like an adaptive-optics PSF.

Of course, redshift-space distortions mean that the true power
spectrum will not be strictly isotropic in practice.  However, the
simulation results described below suggest that the effect of such
deviations are small and do not have a significant impact on the
recovered power spectrum.

In order to demonstrate the effect of this window function on power
spectra, we have sampled realizations of a linear density field
covering the volume of the 2dFGRS data. Averaging the results of this
analysis for 1000 realizations with different phases, and comparing
with the true input power spectrum, we can quantify the effect of the
window function. This is shown in Fig.~\ref{fig:win_pow} for spectra
corresponding to two different cosmologies. The shape of the spectra
are significantly altered, and any oscillations are
damped. Analytically convolving the power spectra with the fit to the
window function given by Equation~\ref{eq:win_fit} provides the same
result.

Because the random catalogue is scaled to match the normalization of
the galaxy catalogue, the average overdensity is artificially set to
zero, forcing $P(0)=0$. This self-normalization results in a deficit
in the measured power spectrum equivalent to subtracting a scaled copy
of the window function, centred on $k=0$. However, this is a very
small effect in the regime of interest because $|W_k|^2$ is a rapidly
decreasing function of $k$ (see Fig.~\ref{fig:win}), and $P(k)$ is
expected to be an increasing function of $k$ for small $k$. The effect
of this self-normalization is shown in Fig.~\ref{fig:win_pow} for
model power spectra.

\begin{figure}
  \setlength{\epsfxsize}{\columnwidth} \centerline{\epsfbox{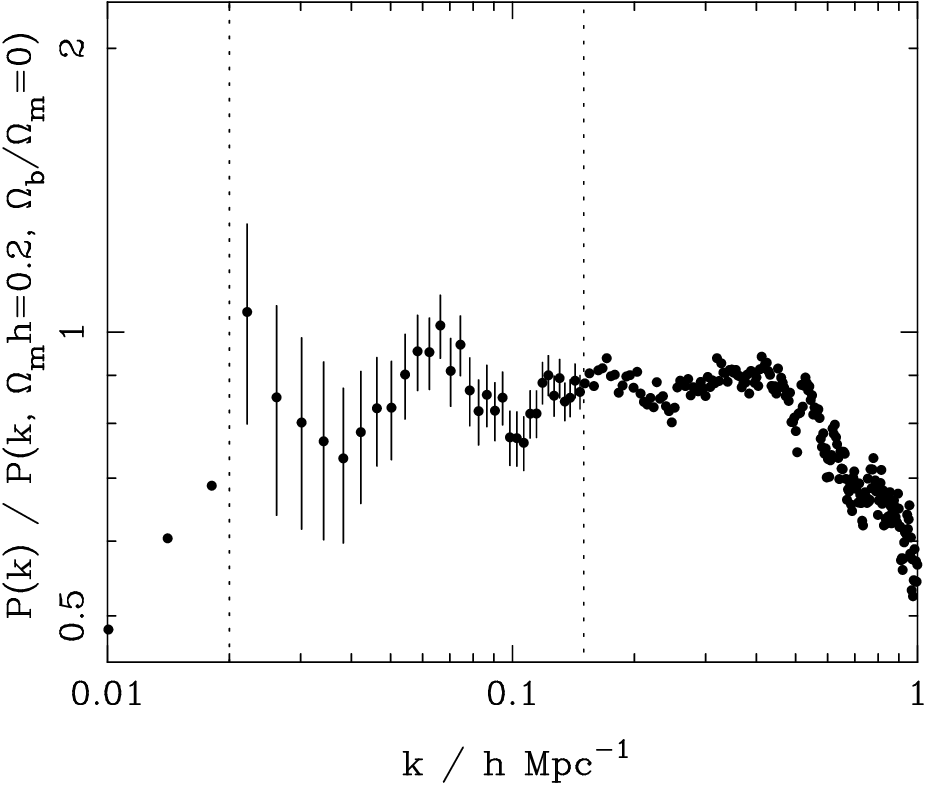}}

  \caption{The 2dFGRS estimate of the redshift-space galaxy power
  spectrum, expressed as the ratio to a linear-theory CDM
  $P(k,\Omega_mh=0.2,\Omega_b/\Omega_m=0)$ power spectrum with $n=1$
  \& $\sigma_8=1$. These data do not estimate the true power spectrum,
  but give the power spectrum convolved with the window function (see
  Section~\ref{sec:window}). Error bars are determined from the
  diagonal elements of the covariance matrix (calculated in
  Section~\ref{sec:cov_mat}), for the $0.02\hompc<k<0.15\hompc$ data
  constrained by the vertical dotted lines. This is the region fitted
  in Section~\ref{sec:fit_mocks}.}

  \label{fig:pk}
\end{figure}

\subsection{Results}  \label{sec:results}

The recovered 2dFGRS power spectrum is presented in Fig.~\ref{fig:pk}.
To highlight features in the spectra presented in this paper, we have
plotted their ratio with a smooth CDM power spectrum
$P(k,\Omega_mh=0.2,\Omega_b/\Omega_m=0)$ that has no baryon
features. We also take a normalization of $\sigma_8=1$ for this
reference model. All of the power spectra used in this paper are
calculated using the transfer function fitting formulae of Eisenstein
\& Hu (1998) and assume a scale-invariant primordial spectrum unless
stated otherwise. We have parametrized the model power spectra
explicitly by $P(k,\Omega_mh,\Omega_b/\Omega_m)$ in order to avoid
confusion with differing definitions of the commonly used shape
parameter $\Gamma$.

The raw results are gratifyingly accurate, with fractional errors in
the power of only $\sim 15\%$ out to $k=0.02\hompc$.  To within about
20\%, the observed spectral shape is that of the
$P(k,\Omega_mh=0.2,\Omega_b/\Omega_m=0)$ reference model between $0.02
\hompc < k < 0.6 \hompc$.  At smaller scales, the finger-of-God
redshift-space smearing is clearly seen to reduce the power.  In many
ways, the most striking features are the suggestions of oscillatory
modulations, with a possible peak at $k\simeq 0.065\hompc$ and
possible troughs at $k\simeq0.035\hompc$ and $k\simeq0.1\hompc$.
However, it is clear that the window function has caused adjacent
power estimates to be closely correlated, so a proper covariance
analysis is required before any significance can be given to these
apparent features.  Given the precision of the basic power estimates,
it is necessary to attain an accurate measure of the systematic
effects listed above that modify the shape of the spectrum.  We
achieve this in the next Section by using mock data.

\begin{figure*}
  \setlength{\epsfxsize}{\textwidth}
  \centerline{\epsfbox{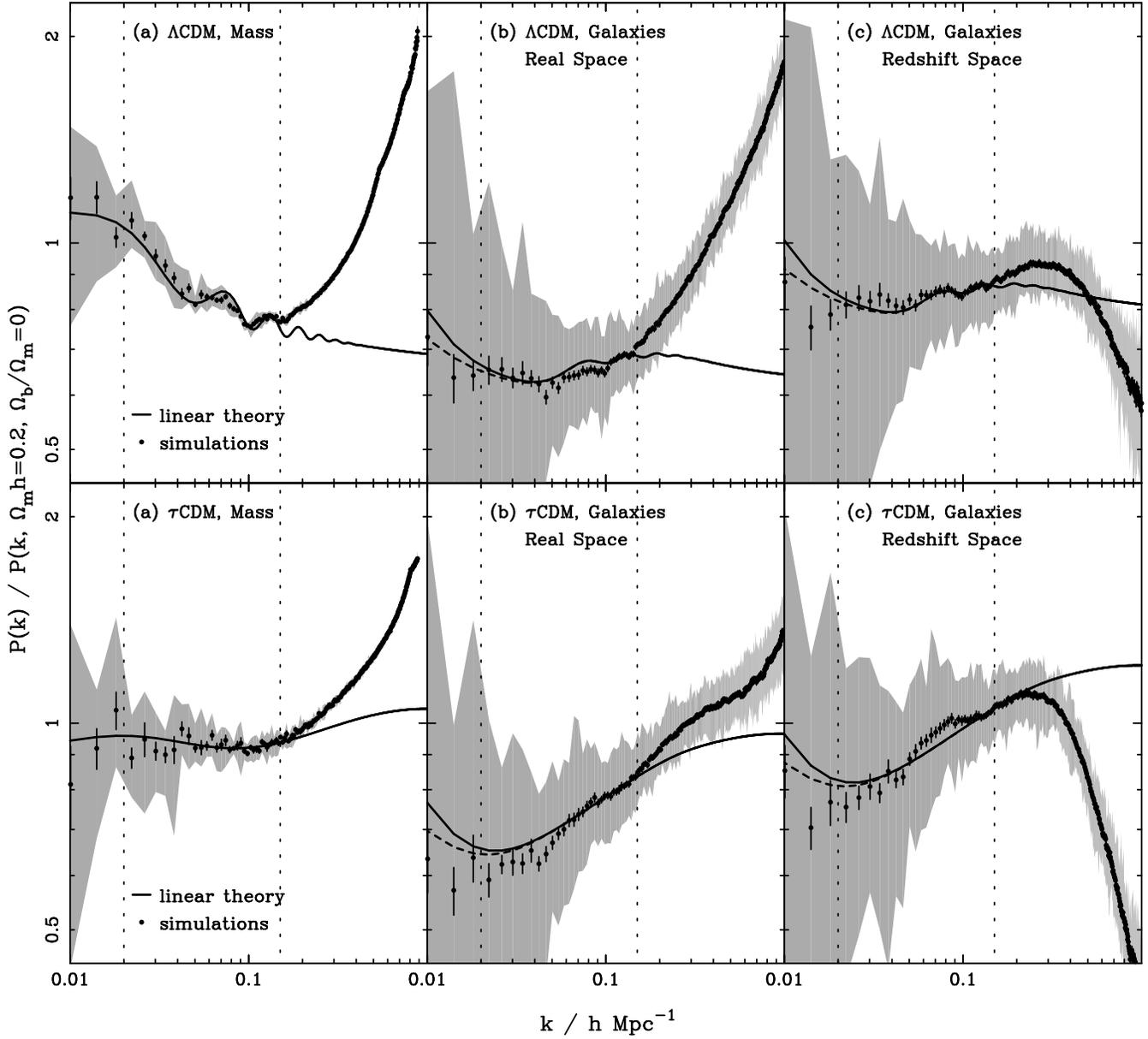}}

  \caption{A comparison of power spectra determined from the results
  of the $\Lambda$CDM and $\tau$CDM Hubble volume simulations divided
  by a $P(k,\Omega_mh=0.2,\Omega_b/\Omega_m=0)$ reference model. (a)
  the average spectrum recovered from the $z=0$ distribution of mass
  in 8 cubes each of side $1500\mpcoh$ ($\Lambda$CDM) or $1000\mpcoh$
  ($\tau$CDM) (solid circles with 1$\sigma$ errors: the extremal data
  bound the shaded area), compared to the scaled input power spectrum
  (solid line). Here, the difference between the two is dominated by
  the non-linear collapse of small-scale structures at
  $k>0.15\hompc$. The $\tau$CDM mass power spectrum is compared to a
  reference model with $\sigma_8=0.6$ (rather than $\sigma_8=1.0$) in
  order that it has approximately the same normalization as the other
  data. (b) the average recovered power spectrum from 35
  ($\Lambda$CDM) and 30 ($\tau$CDM) real-space galaxy catalogues
  sampled from the simulations to match the 2dFGRS window function
  (solid circles with 1$\sigma$ errors: the extremal data bound the
  shaded area). These are compared to the linear spectrum convolved
  with the analytic approximation to the window function given by
  Equation~\ref{eq:win_fit} (solid line). The difference between these
  two is dominated by both bias (used to create the galaxy catalogue)
  and non-linear effects. (c) the average recovered power spectra from
  35 ($\Lambda$CDM) and 30 ($\tau$CDM) redshift-space galaxy
  catalogues designed to fully mimic the 2dFGRS redshift-space sample
  (solid circles with 1$\sigma$ errors: the extremal data bound the
  shaded area), and the linear spectra convolved with the window
  function (solid line). Here, the differences are due to all of the
  factors listed in Section~\ref{sec:FKP_method}. The dashed lines in
  panels (b)~\&~(c) show the power spectrum expected from data in
  which the average over-density is artificially set to zero (see
  Section~\ref{sec:window} for details). The $0.02\hompc<k<0.15\hompc$
  region fitted in Section~\ref{sec:fit_mocks} is delineated by the
  vertical dotted lines. In calculating the error in the average power
  spectra measured from the simulations, we have assumed that the
  samples are independent. This is not precisely true, and the errors
  plotted therefore underestimate the true error.}

  \label{fig:mocks}
\end{figure*}

\section{Tests on mock data}
\label{sec:region}

Redshift-space and non-linear effects mean that the shape of the
recovered power spectrum gives information about the linear power
spectrum only at small values of $k$. The easiest way to model both
these effects is via numerical simulation. Using an empirically
motivated biasing scheme, it is possible to place galaxies within
$N$-body simulations and provide mock catalogues designed to mimic the
2dFGRS catalogue for different cosmological models (Cole \etal\ 1998;
Baugh \etal\ 2001). In this Section we use these mock catalogues to
demonstrate that for $k<0.15\hompc$\ the only important effect on the
power spectrum is the convolution with the window function.

Fig.~\ref{fig:mocks} shows a collection of power spectra calculated
from samples drawn from the $\Lambda$CDM and $\tau$CDM Hubble volume
simulations (see Cole \etal\ 1998, Baugh \etal\ 2001 and {\tt
http://star-www.dur.ac.uk/{\tt\char'176}cole/mocks/main.html} for
details). The large catalogues constructed in this work were
re-sampled depending on angular position and redshift to match the
correct window function for the 2dFGRS data. Power spectra were
calculated exactly as for the 2dFGRS data. Fig.~\ref{fig:mocks}
illustrates the factors that transform the linear mass power spectrum
into the non-linear galaxy spectrum. Panel (a) shows the power
spectrum of the mass at $z=0$. This demonstrates the increase in power
on small scales caused by the collapse of halos. Panel (b) differs in
that we now have to consider the effect of bias (artificially added to
the simulation), and the window function. In panel (c), we analyze
samples designed to mimic the 2dFGRS data as closely as possible,
including redshift-space effects: the finger-of-God effect that
decreases small-scale power, and the Kaiser effect that enhances the
power. The redshift-space and non-linear effects cancel to some extent
and give approximately the correct level of $P(k)$ out to
$k\sim0.5\hompc$. However, the shape of the power spectrum is altered
for $k\simgt0.2\hompc$.

We will therefore assume that, at $k<0.15\hompc$, redshift space
distortions and non-linear effects have an insignificant effect on the
shape of $P(k)$. Allowing the normalization to vary removes any
large-scale constant bias; the bias is not expected theoretically to
vary significantly with $k$ on these large scales (Kauffmann, Nusser
\& Steinmetz 1997; Benson \etal\ 2000).  In the future, measurements
of $\beta\equiv\Omega_m^{0.6}/b$ from redshift-space distortions as a
function of scale will test directly the degree to which this is true
(see Peacock \etal\ 2001 for the first 2dFGRS results on
redshift-space distortions).  This region of the power spectrum
therefore directly provides information about the shape of the linear
power spectrum, and can be used with models of the transfer function
to provide constraints on cosmological parameters. Although this
conclusion has only been justified here for two specific assumed
models, we have performed similar tests on a wider range of models. In
Table 1 below, we show explicitly how the results depend on the range
of wavenumber considered.

\section{estimating the covariance matrix}  \label{sec:cov_mat}

The $P(k)$ data points in Fig.~\ref{fig:pk} are not independent, and
correlations extend across finite regions of $k$-space. This is
predominantly caused by the convolution with the window function,
although there is also a contribution from non-linear effects (Meiksin
\& White 1999; Scoccimarro, Zaldarriaga \& Hui 1999) and
redshift-space effects. Particular care must therefore be taken in
interpreting `wiggles' in the power spectrum as significant
features. In order to quantify these correlations, we have estimated
the covariance matrix for the data points with $k<0.15\hompc$.

A large number of independent realizations of $P(k)$ are required in
order to have sufficient signal-to-noise in the covariance matrix. It
would be too time-consuming to perform separate numerical simulations
for each data set. Instead, we have created 1000 realizations of a
Gaussian random field on a $256\times256\times128$ grid covering the
region of the 2dFGRS survey. For $k<0.15\hompc$, using a smaller grid
than that used for the 2dFGRS data does not significantly affect the
result, and reduces the computational burden. The resulting power
spectra, determined as for the 2dFGRS data, were used in
Section~\ref{sec:window} to demonstrate the effect of convolving
$P(k)$ by the window function. These realizations provide an estimate
of the cosmic variance within the 2dFGRS volume. The contribution from
shot noise has been calculated by analysing similar Monte-Carlo
realizations, and has been included in our estimate of the covariance
matrix.

Estimating the covariance matrix in this way does not take into
account non-linear and redshift-space effects, which add to the
covariances. However, these effects should be small over the region of
$k$-space we are fitting. As a test of this we have estimated the
covariance matrix using 10 catalogues drawn from $\Omega_mh=0.25$
$\Omega_b/\Omega_m=0$ CDM simulations with different phases by Cole
\etal\ (1998). These catalogues were calculated using Cole \etal\
(1998) bias model~1. The correlations calculated over the $k$-space
region $0.02\hompc<k<0.15\hompc$\ were similar in scale to those
calculated from the Gaussian fields. However, we find that the errors
in $P(k)$ calculated from numerical simulations are $16$\% larger than
those determined from Gaussian simulations, although there is no
evidence for a change in shape of the diagonal elements of the
covariance matrix for $k<0.15\hompc$. We do see a change at
$k>0.15\hompc$ consistent with non-linear and redshift-space effects,
which are expected to be an increasing function of $k$. In this work
we adopt the conservative approach and renormalize the covariance
matrix calculated from the Gaussian realisations to match the
normalisation of the numerical simulations, whilst keeping the
correlation matrix the same. This renormalization does not
significantly affect the primary results of this paper: the derived
best-fit parameters are the same with or without this renormalization.

\section{fitting to the power spectrum}  \label{sec:fit}

\subsection{Model parameters}

Model power spectra for different cosmologies have been created to
compare with the 2dFGRS data using the transfer function fitting
formulae of Eisenstein \& Hu (1998). These formulae are essentially
perfect in the regime of interest, and show that the effect of baryons
on the power spectrum is not well approximated by a change in
$\Omega_mh$. These differences are important in order to be able to
fit the 2dFGRS data with both parameters. The shape of the power
spectrum is primarily dependent on $\Omega_mh$, and only weakly
dependent on $h$. We have therefore chosen to fit $\Omega_mh$ rather
than $\Omega_m$. Similarly, the strength of baryon oscillations depend
primarily on $\Omega_b/\Omega_m$. We have therefore calculated model
spectra for cosmologies chosen on a $40\times40\times40$ grid in
$(\Omega_mh, \Omega_b/\Omega_m , h)$ space, covering
\begin{eqnarray}
  0.1 \quad < & \Omega_mh         & < \quad 0.8, \nonumber \\
  0.0 \quad < & \Omega_b/\Omega_m & < \quad 0.5, \\
  0.4 \quad < & h                 & < \quad 0.9. \nonumber
\end{eqnarray} 
These limits effectively act as uniform prior probability densities
for the parameters. A scale-invariant $n=1$ primordial spectrum was
assumed initially, following results from CMB analysis (e.g. Jaffe
\etal\ 2000). We have numerically convolved these power spectra on the
scales $0.02\hompc<k<0.15\hompc$ using the window function fitting
formula presented in Section~\ref{sec:window}. The normalization of
the model spectra was allowed to vary to account for an unknown
large-scale linear bias.

Following analysis of the data using these model spectra, we have also
created models on a finer $40\times40\times40$ grid covering
$0.1<\Omega_mh<0.3$, $0.0<\Omega_b/\Omega_m<0.4$ and $0.4<h<0.9$ in order
to further constrain the fit in this region of parameter space.

We emphasise that $\Omega_m$ is the total matter density parameter,
i.e. $\Omega_m=\Omega_{\rm cdm}+\Omega_b$. Our results are not
strongly dependent on $\Omega_v$.

\subsection{Results}  \label{sec:fit_results}

The likelihood of each model has been estimated using a covariance
matrix calculated from Gaussian realizations of linear density fields
as in Section~\ref{sec:cov_mat}. Results presented in this Section
were calculated using the covariance matrix estimated by the method
described in Section~\ref{sec:cov_mat}, for which $\chi^2_{\rm
min}=27.5$, given an expected value of $28$. The best fit power
spectrum parameters are only weakly dependent on this choice: we have
tested a number of covariance matrices calculated for different input
power spectra and find $\chi^2_{\rm min}$ close to the expected value
for all of them. In fact, we used an iterative procedure leading to
this choice of covariance matrix. Initially we adopted a
$\Omega_mh=0.25$, $\Omega_b/\Omega_m=0$ power spectrum, and then
adopted the values $\Omega_mh=0.2$ and $\Omega_b/\Omega_m=0.16$ which
are close to the best fit values determined with this covariance
matrix.

\begin{figure}
  \setlength{\epsfxsize}{\columnwidth}
  \centerline{\epsfbox{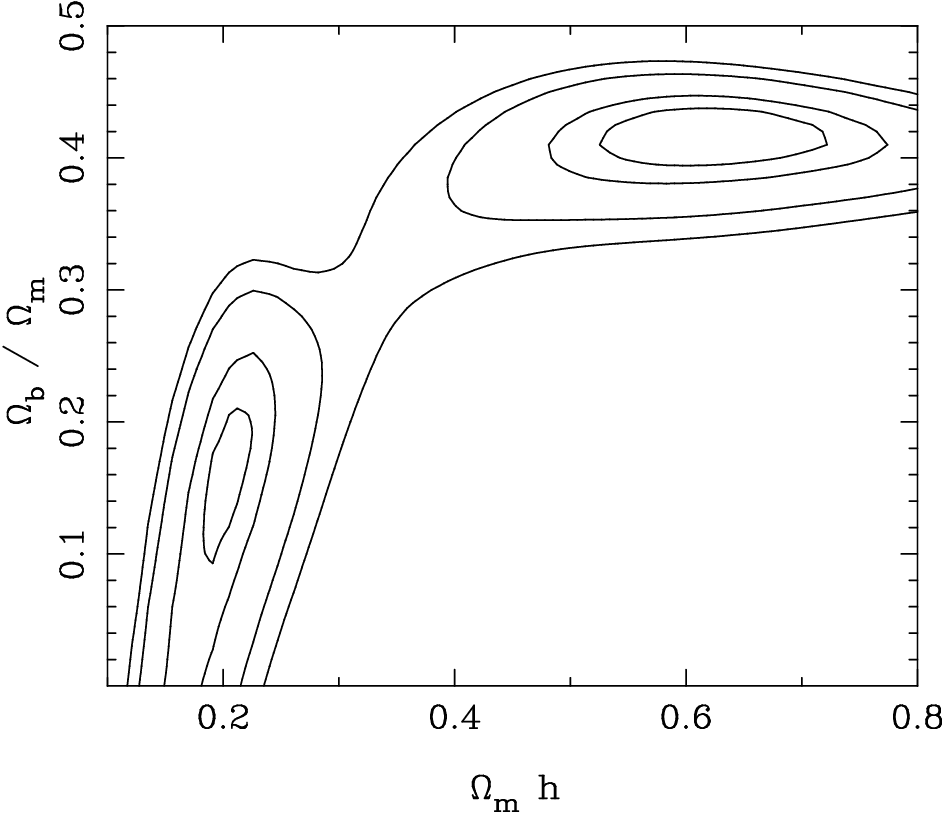}} 

  \caption{Likelihood surfaces for the best fit linear power spectrum
  over the region $0.02\hompc<k<0.15\hompc$. The normalization is a
  free parameter to account for the unknown large-scale
  biasing. Contours are plotted at the usual positions for
  one-parameter confidence of 68\%, and two-parameter confidence of
  68\%, 95\% and 99\% (i.e. $-2\ln({\cal L}/{\cal L_{\rm max}}) = 1,
  2.3, 6.0, 9.2$). We have marginalized over the missing free
  parameters ($h$ and the power spectrum amplitude) by integrating
  under the likelihood surface.}

  \label{fig:powfit1}
\end{figure}

The likelihood contours in $\Omega_b/\Omega_m$ versus $\Omega_mh$ for
this fit are shown in Fig.~\ref{fig:powfit1}. At each point in this
surface we have marginalized by integrating the likelihood surface
over the two free parameters, $h$ and the power spectrum
amplitude. The result is not significantly altered if instead, the
modal, or Maximum Likelihood points in the plane corresponding to
power spectrum amplitude and $h$ were chosen. The likelihood function
is also dependent on the covariance matrix (which should be allowed to
vary with cosmology), although the consistency of result from
covariance matrices calculated for different cosmologies shows that
this dependence is negligibly small. Thus ${\cal
L}\propto\exp(-\chi^2/2)$ in practice.

Fig.~\ref{fig:powfit1} shows that there is a degeneracy between
$\Omega_mh$ and the baryonic fraction $\Omega_b/\Omega_m$. However, there
are two local maxima in the likelihood, one with $\Omega_mh \simeq 0.2$
and $\sim 20\%$ baryons, plus a secondary solution $\Omega_mh \simeq 0.6$
and $\sim 40\%$ baryons.

\begin{figure}
  \setlength{\epsfxsize}{\columnwidth}
  \centerline{\epsfbox{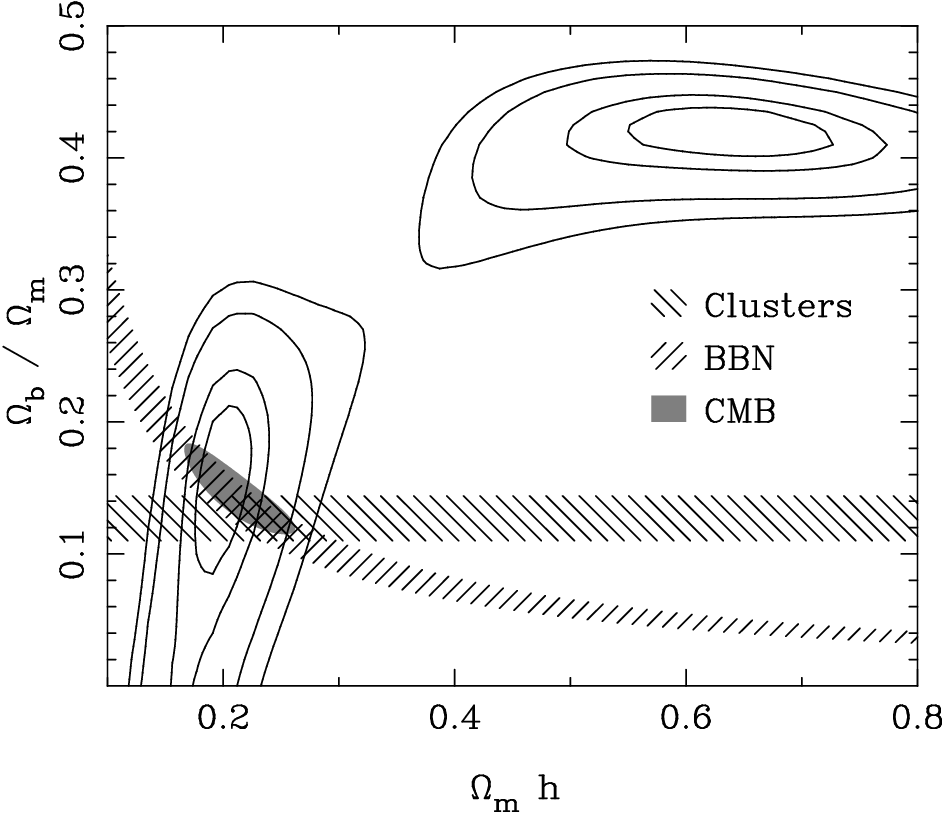}} 

  \caption{Likelihood surfaces for the best fit linear power spectrum
  over the region $0.02\hompc<k<0.15\hompc$, as in
  Fig.~\ref{fig:powfit1}, but now adding a prior on $h$: $h=0.7\pm
  10\%$. This tightens the constraints. This result is compared to
  estimates from X-ray cluster analysis (Evrard 1997), big-bang
  nucleosynthesis (O'Meara \etal\ 2001) and recent CMB results
  (Netterfield \etal\ 2001; Pryke \etal\ 2001). Note that we have
  plotted the CMB result following the reasonable approximation that
  $\Omega_bh^2$ and $\Omega_{\rm cdm}h^2$ were independently
  determined by each of these analyses.}

  \label{fig:powfit2}
\end{figure}

Assuming a uniform prior for $h$ over a factor of 2 is arguably
over-cautious, and we have therefore multiplied by a Gaussian prior
$h=0.7\pm 10\%$ in Fig.~\ref{fig:powfit2}. This corresponds to
multiplying by the likelihood from external constraints such as the
HST key project (Freedman \etal\ 2000). The effect is to tighten the
contours around the above two models. The low-density model now
becomes approximately
\begin{equation}
  \Omega_m h = 0.20 \pm 0.03; \quad\quad \Omega_b/\Omega_m = 0.15 \pm 0.07.
\end{equation}
The errors quoted are rms errors, and have been calculated by
integrating over the branch of solutions of interest. Analysing mock
catalogues drawn from the $\Lambda$CDM Hubble volume simulation
produces similar rms errors, and shows that, for each parameter, the
interval defined by the appropriate error is close to a 68\%
confidence interval (see Section~\ref{sec:fit_mocks}).

\begin{table}
  \centering \begin{tabular}{cccccc} \hline
  \multicolumn{2}{c}{$k$ range / $\hompc$} & n & assumed &
  \multicolumn{2}{c}{ML parameters} \\
  min & max & & $\Omega_m$ for $r(z)$ & $\Omega_mh$ & $\Omega_b/\Omega_m$ \\
  \hline

  0.02  & 0.15 & 1.0 & 1.0 & 0.23$\pm$0.03 & 0.18$\pm$0.07 \\  % checked
  0.02  & 0.15 & 1.0 & 0.4 & 0.20$\pm$0.03 & 0.16$\pm$0.07 \\  % checked
  0.02  & 0.15 & 0.9 & 0.3 & 0.22$\pm$0.03 & 0.12$\pm$0.07 \\  % checked
  0.02  & 0.15 & 1.1 & 0.3 & 0.18$\pm$0.03 & 0.19$\pm$0.07 \\  % checked
  0.015 & 0.15 & 1.0 & 0.3 & 0.20$\pm$0.03 & 0.14$\pm$0.07 \\  % checked
  0.03  & 0.15 & 1.0 & 0.3 & 0.20$\pm$0.03 & 0.15$\pm$0.07 \\  % checked
  0.02  & 0.10 & 1.0 & 0.3 & 0.17$\pm$0.04 & 0.18$\pm$0.08 \\  % checked
  0.02  & 0.12 & 1.0 & 0.3 & 0.18$\pm$0.03 & 0.17$\pm$0.07 \\  % checked
  0.02  & 0.15 & 1.0 & 0.3 & 0.20$\pm$0.03 & 0.15$\pm$0.07 \\  % checked
  \hline
  \end{tabular}

  \caption{Maximum Likelihood $\Omega_mh$ and $\Omega_b/\Omega_m$
  parameters for fits to the 2dFGRS power spectrum, varying the range
  of $k$-space fitted, the power-law index $n$ of the primordial
  spectrum, and the matter density of the flat cosmology assumed to
  estimate the comoving distance to each galaxy. All of these fits
  used a covariance matrix calculated from Gaussian realizations of a
  $\Omega_mh=0.2$, $\Omega_b/\Omega_m=0.16$ CDM power spectrum.}
  \label{tab:fit_data}
\end{table}

\begin{figure}
  \setlength{\epsfxsize}{\columnwidth} 
  \centerline{\epsfbox{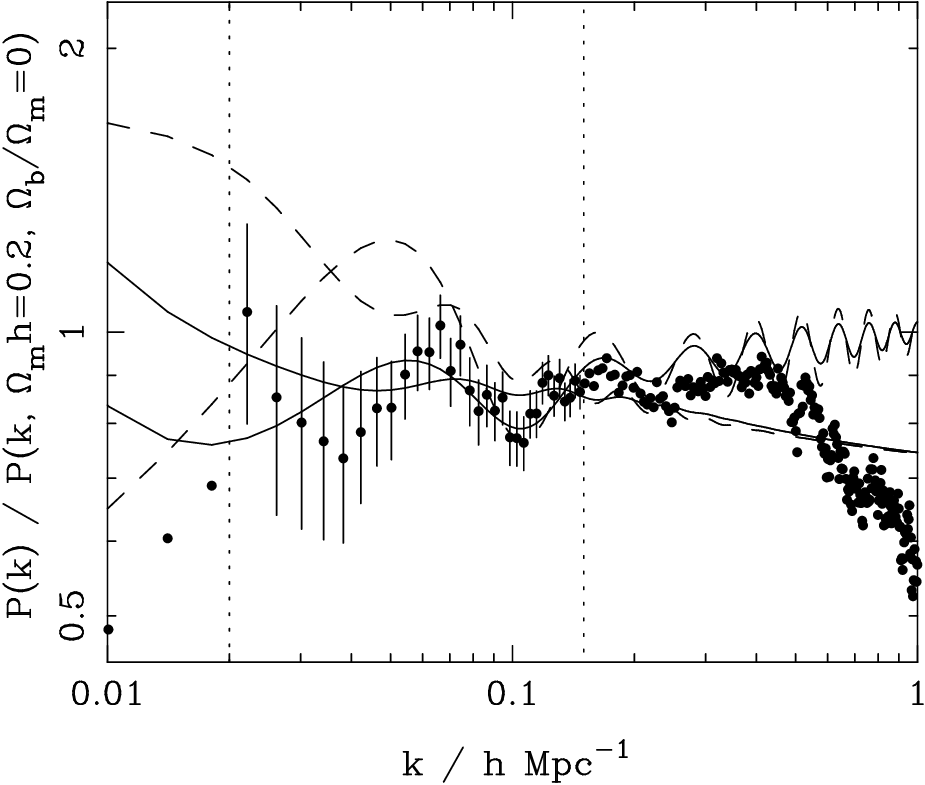}}

  \caption{The 2dFGRS data compared with the two preferred models from
  the Maximum Likelihood fits convolved with the window function
  (solid lines). Error bars show the diagonal elements of the
  covariance matrix, for the fitted data that lie between the dotted
  vertical lines. The unconvolved models are also shown (dashed
  lines). The $\Omega_m h \simeq 0.6$, $\Omega_b/\Omega_m=0.42$,
  $h=0.7$ model has the higher bump at $k\simeq 0.05\hompc$. The
  smoother $\Omega_m h \simeq 0.20$, $\Omega_b/\Omega_m=0.15$, $h=0.7$
  model is a good fit to the data because of the overall shape.}

  \label{fig:models}
\end{figure}

The 2dFGRS data are compared to the best-fit linear power spectra
convolved with the window function in Fig.~\ref{fig:models}. This
shows where the two branches of solutions come from: the low-density
model fits the overall shape of the spectrum with relatively small
`wiggles', while the solution at $\Omega_m h \simeq 0.6$ provides a
better fit to the bump at $k\simeq 0.065\hompc$, but fits the overall
shape less well.

\subsection{Robustness of the fit}  \label{sec:fit_testing}

We have tried varying the range of $k$ for the fit, the assumed
power-law index of the primordial fluctuations, and the assumed
geometry.  The best fit $\Omega_mh$ and $\Omega_b/\Omega_m$ for a
variety of assumptions are presented in Table~\ref{tab:fit_data} along
with approximate errors. $h=0.7\pm 10\%$ was assumed for this
analysis. The shape of the likelihood surfaces and the position of the
minimum recovered from each of these fits are similar, and the ML
values generally change by $\ll 1 \sigma$. The main effect of changing
assumptions is in how rapidly the likelihood falls away from the ML
point.

Perhaps the main point to emphasize here is that the results are not
greatly sensitive to the assumed tilt of the primordial spectrum. We
have used the CMB results to motivate the choice of $n=1$, but it is
clear that very substantial tilts are required to alter our
conclusions significantly: $n\simeq 0.8$ would be required for the
zero baryon model to become an acceptable fit, within 1$\sigma$ of
the preferred model.

\subsection{Fitting to mock data}  \label{sec:fit_mocks}

\begin{figure}
  \setlength{\epsfxsize}{\columnwidth}
  \centerline{\epsfbox{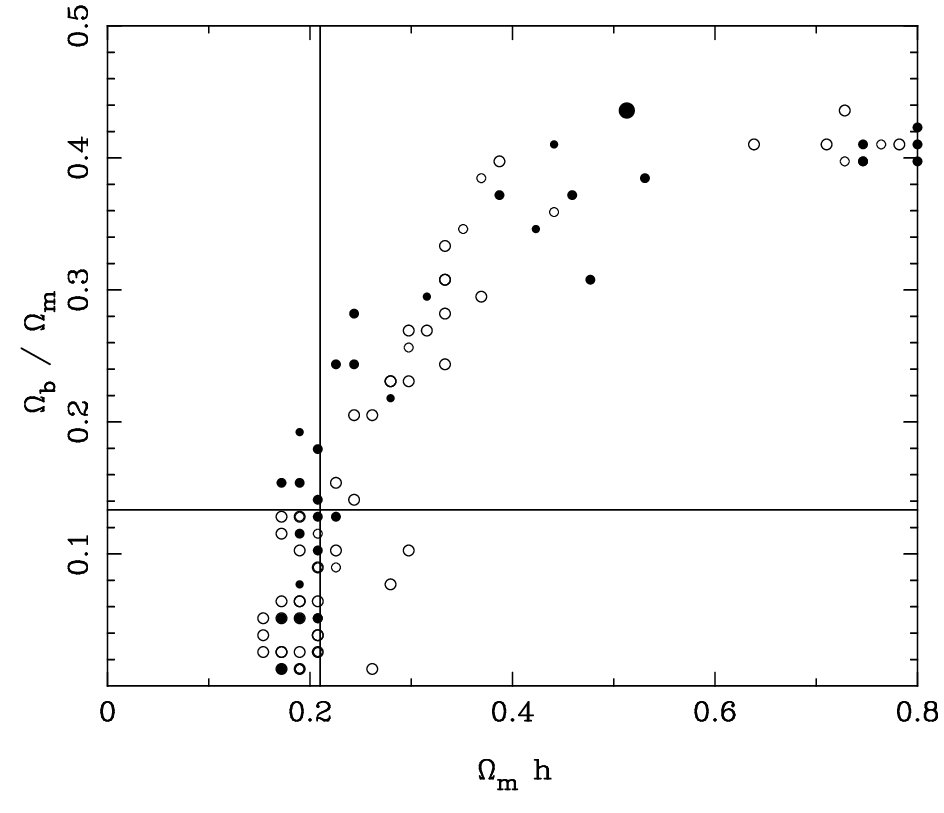}} 

  \caption{2D likelihood maxima in the $(\Omega_mh,
  \Omega_b/\Omega_m)$ plane calculated from 35 redshift-space
  catalogues drawn from the $\Lambda$CDM Hubble volume simulation. We
  have marginalized (by integrating under the likelihood) over $h$ and
  $P(k)$ amplitude assuming a Gaussian prior $h=0.7\pm0.07$ and a
  uniform prior for the amplitude. Primary maxima are given by the
  solid circles, other maxima with $-2\ln({\cal L}/{\cal L_{\rm
  max}})<9.2$ (of which there were $\sim1.7$ for each catalogue) by
  the open circles. Primary maxima for which there was no secondary
  maximum are shown by the large solid circles. The locus of recovered
  parameters has a similar pattern to that of power spectra with
  similar shape over $0.02\hompc<k<0.15\hompc$, and is also similar to
  that seen in the likelihood contours deduced from the 2dFGRS
  data. Solid lines mark the input parameters for the simulation.}

  \label{fig:fit_mocks}
\end{figure}

Using the technique described in Section~\ref{sec:fit_results}, we
have tried to recover the input parameters of the $\Lambda$CDM
simulation, $\Omega_mh=0.21$ and $\Omega_b/\Omega_m=0.13$, from 35
redshift-space galaxy catalogues drawn from this simulation to mimic
the 2dFGRS data. Having calculated power spectra for the catalogues
(presented in panel c of Fig.~\ref{fig:mocks}), we have fitted the
data with models using a covariance matrix calculated as in
Section~\ref{sec:cov_mat}. The Maximum Likelihood (and secondary
maxima if present) $\Omega_mh$ and $\Omega_b/\Omega_m$ parameters
recovered from these catalogues are presented in
Fig.~\ref{fig:fit_mocks}. There is a degeneracy between recovered
parameters: the data trace a pattern similar to that of power spectra
with similar shape in the region $0.02\hompc<k<0.15\hompc$. This
pattern is similar to that determined from the 2dFGRS data
(Fig.~\ref{fig:powfit2}). The degeneracy between $\Omega_mh$ and
$\Omega_b/\Omega_m$ is weakly broken with the models favouring
approximately the correct baryon fraction. For each maximum within the
region $0.1<\Omega_mh<0.3$ \& $0.0<\Omega_b/\Omega_m<0.4$, we have
integrated over the likelihood and find rms values similar to those
obtained from the 2dFGRS data. Of the 35 catalogues modelled, 27
($77\%$) have likelihood maxima within 1~rms of the true $\Omega_mh$
value, (i.e. they have likelihood maxima with
$0.18<\Omega_mh<0.24$). We also find that 24 ($69\%$) have likelihood
maxima within 1~rms of the true baryon fraction (i.e. they have
likelihood maxima with $0.06<\Omega_b/\Omega_m<0.20$). This gives us
confidence that the quoted errors for the best-fit parameters derived
from the 2dFGRS data are sound.

Combining the likelihood surfaces calculated from the 13
non-overlapping catalogues results in best-fit parameters
$\Omega_mh=0.20\pm0.15$ and $\Omega_b/\Omega_m=0.10\pm0.05$, with
$-2\ln({\cal L_{\rm true}}/{\cal L_{\rm max}})<1.0$ for the combined
likelihood. This offset is statistically acceptable, and suggests
that, if there is a systematic bias in determining $\Omega_mh$ and
$\Omega_b/\Omega_m$ from these catalogues, it is at a level well below
the errors on the recovered parameters from any single catalogue.

\section{Conclusions}  \label{sec:conc}

We have shown that the present 2dFGRS data allow the galaxy power
spectrum to be measured to high accuracy (10--15\% rms) over about a
decade in scale at $k<0.15\hompc$.  We have carried out a range of
tests for systematics in the analysis and a detailed comparison with
realistic mock samples.  As a result, we are confident that the 2dFGRS
result can be interpreted as giving the shape of the linear-theory
matter power spectrum on these large scales, and that the statistical
errors and covariances between the data points are known.

By fitting our results to the space of CDM models, we have been able
to reach a number of interesting conclusions regarding the matter
content of the universe:

\japitem{(1)}The power spectrum is close in shape to that of a
$\Omega_mh=0.2$ model, to a tolerance of about 20\%.

\japitem{(2)}Nevertheless, there is sufficient structure in the $P(k)$
data that the degeneracy between $\Omega_b/\Omega_m$ and $\Omega_mh$
is weakly broken. The two local likelihood maxima have $(\Omega_mh,
\Omega_b/\Omega_m) \simeq (0.2,0.15)$ and $(0.6,0.4)$ respectively.

\japitem{(3)}The evidence for detection of baryon oscillations in the
power spectrum is presently modest, with a likelihood ratio of
approximately $2.7$ between the favoured model and the best
zero-baryon model. Conversely, a large baryon fraction can be very
strongly excluded: $\Omega_b/\Omega_m < 0.28$ at 95\% confidence,
provided $\Omega_mh < 0.4$.

\japitem{(4)}These conclusions do not depend strongly on the value of
$h$, although they do depend on the tilt of the primordial spectrum,
with $n\simeq 0.8$ being required to make a zero-baryon model an
acceptable fit to the data.

\japitem{(5)}The sensitivity to tilt emphasizes that the baryon signal
comes in good part from the overall shape of the spectrum. Although
the eye is struck by a single sharp `spike' at $k\simeq 0.065\hompc$,
the correlated nature of the errors in the $P(k)$ estimate means that
such features tend not to be significant in isolation. We note that
the convolving effects of the window would require a very substantial
spike in the true power in order to match our data exactly. Such
over-fitting is not possible within the compass of conventional
models, and the conservative conclusion is that the apparent spike is
probably enhanced by correlated noise.  A proper statistical treatment
is essential in such cases.

\enditem 
It is interesting to compare these conclusions with other
constraints. Averaging the results of Netterfield \etal\ 2001 and
Pryke \etal\ 2001, the current CMB data require $\Omega_m h^2=0.15 \pm
0.03$, $\Omega_b h^2 = 0.0215 \pm 0.0025$, together with a
power-spectrum index of $n=0.97 \pm 0.06$, on the assumption of pure
scalar fluctuations.  If we take $h=0.7\pm 10\%$, this gives
\begin{equation}
  \Omega_m h=0.21 \pm 0.05;\quad\quad\Omega_b/\Omega_m=0.14\pm 0.03,
\end{equation}
in remarkably good agreement with the estimate from the 2dFGRS
\begin{equation}
  \Omega_m h = 0.20 \pm 0.03; \quad\quad \Omega_b/\Omega_m = 0.15 \pm 0.07.
\end{equation}
Latest estimates of the Deuterium to Hydrogen ratio in QSO spectra
combined with big-bang nucleosynthesis theory predict $\Omega_bh^2 =
0.0205\pm 0.0018$ (O'Meara \etal\ 2001), in agreement with the latest
CMB results. The confidence interval estimated from the 2dFGRS power
spectrum overlaps both regions. X-ray cluster analysis predicts a
baryon fraction $\Omega_b/\Omega_m=0.127\pm0.017$ (Evrard 1997) which
is again within $1\sigma$ of our preferred value.

The above limits are all shown on Fig.~\ref{fig:powfit2}, and paint a
picture of impressive consistency: it appears that we live in a
universe that has $\Omega_m\simeq0.3$ with a baryon fraction of
approximately $15\%$. The precision of this statement will improve
greatly with completion of the 2dFGRS. Doubling the sample size will
improve the errors on the baryon fraction by much more than $\sim
\sqrt{2}$.  The window function will be more compact, so the
signatures of baryon oscillations should be seen very clearly -- or we
will see that some non-standard alternative is required.

\section*{ACKNOWLEDGEMENTS}

The 2dF Galaxy Redshift Survey was made possible through the dedicated
efforts of the staff of the Anglo-Australian Observatory, both in
creating the 2dF instrument and in supporting it on the telescope.


\begin{thebibliography}{}
  \bibitem[\protect\citename{Baugh \& Efstathiou }1993]{baugh} 
    Baugh C.M., Efstathiou G., 1993, MNRAS, 265, 145
  \bibitem[\protect\citename{Baugh \etal }2001]{baugh2001} 
    Baugh C.M., \etal, 2001, in preparation
  \bibitem[\protect\citename{Benson \etal\ }2000]{benson} 
    Benson A.J., Cole S., Frenk C.S., Baugh C.M., Lacey C.G., 2000, MNRAS, 311, 793
  \bibitem[\protect\citename{Cole \etal\ }1998]{cole} 
    Cole S., Hatton S., Weinberg D.H., Frenk C.S., 1998, MNRAS, 300, 945
  \bibitem[\protect\citename{Colless }1999]{colless} 
    Colless M.M., 1999, Phil. Trans. Roy. Soc. Lond. A, 357, 105
  \bibitem[\protect\citename{Efstathiou \& Moody }2000]{em} 
    Efstathiou G., Moody S.J., 2000, astro-ph/0010478
  \bibitem[\protect\citename{Eisenstein \& Hu }1998]{eh98} 
    Eisenstein D.J., Hu W., 1998, ApJ, 496, 605
  \bibitem[\protect\citename{Evrard }1997]{evrard} 
    Evrard A.E., 1997, MNRAS, 292, 289
  \bibitem[\protect\citename{Feldman, Kaiser \& Peacock }1994]{fkp} 
    Feldman H.A., Kaiser N., Peacock J.A., 1994, ApJ, 426, 23
  \bibitem[\protect\citename{Freedman \etal\ }2000]{freedman} 
    Freedman W.L., \etal, 2000, astro-ph/0012376
  \bibitem[\protect\citename{Huchra \etal\ }1990]{huchra} 
    Huchra J.P., Geller M.J., de Lapparant V., Corwin H.G., 1990, ApJS, 72, 433
  \bibitem[\protect\citename{Jaffe \etal\ }2000]{jaffe} 
    Jaffe A., \etal, 2000, Phys. Rev. Lett., 86, 3475
  \bibitem[\protect\citename{Kaiser}1987]{kaiser} 
    Kaiser N., 1987, MNRAS, 227, 1
  \bibitem[\protect\citename{Kauffmann \etal\ }1997]{kauffmann} 
    Kauffmann G., Nusser A., Steinmetz M., MNRAS, 286, 795
  \bibitem[\protect\citename{Lewis \etal\ }2000]{lewis} 
    Lewis I., \etal, 2000, MNRAS submitted
  \bibitem[\protect\citename{Maddox \etal\ }1990a]{apmcounts} 
    Maddox S.J., Efstathiou G., Sutherland W.J., Loveday J., Peterson B.A., 1990a, MNRAS, 242, 43P
  \bibitem[\protect\citename{Maddox \etal\ }1990b]{apmI} 
    Maddox S.J., Efstathiou G., Sutherland W.J., Loveday J., 1990b, MNRAS, 243, 692
  \bibitem[\protect\citename{Maddox \etal\ }1990c]{apmII} 
    Maddox S.J., Efstathiou G., Sutherland W.J., 1990c, MNRAS, 246, 433
  \bibitem[\protect\citename{Maddox \etal\ }1996]{maddox96} 
    Maddox S.J., Efstathiou G., Sutherland W.J., 1996, MNRAS, 283, 1227
  \bibitem[\protect\citename{Meiksin, White \& Peacock }1999]{meiksin99} 
    Meiksin A.A., White M., 1999, MNRAS, 308, 1179
  \bibitem[\protect\citename{Netterfield \etal\ }2001]{netterfield} 
    Netterfield C.B., \etal, 2001, astro-ph/0104460
  \bibitem[\protect\citename{O'Meara \etal\ }2001]{omera} 
    O'Meara J.M., Tytler D., Kirkman D., Suzuki N., Prochaska J.X.,
    Lubin D., Wolfe A.M., 2001, astro-ph/0011179 
  \bibitem[\protect\citename{Peacock \etal\ }2001]{peacock} 
    Peacock J.A., \etal, 2001, Nature, 410, 169
  \bibitem[\protect\citename{Pryke \etal\ }2001]{pryke} 
    Pryke C., \etal, 2001, astro-ph/0104490
  \bibitem[\protect\citename{Saunders \etal\ }2000]{saunders} 
    Saunders W., \etal, 2000, MNRAS, 317, 55 
  \bibitem[\protect\citename{Schlegel \etal\ }1998]{schlegel} 
    Schlegel D.J., Finkbeiner D.P., Davis M., 1998, ApJ, 500, 525
  \bibitem[\protect\citename{Shectman \etal\ }1996]{shectman} 
    Shectman S.A., Landy S.D., Oemler A., Tucker D.L., Lin H.,
    Kirshner R.P., Schechter P.L., 1996, ApJ, 470, 172
  \bibitem[\protect\citename{Scoccimarro \etal\ }1999]{scoccimarro} 
    Scoccimarro R., Zaldarriaga M., Hui L., 1999, ApJ, 527, 1
\end{thebibliography}
\end{document}